\documentclass[aip,apl,superscriptaddress,citeautoscript,preprint]{revtex4-1}
\usepackage{graphicx}
\newcommand{\YBCO}{YBa$_2$Cu$_3$O$_{6+x}$}

\begin{document}

\title{Magnetic force microscopy measurement of the penetration depth in superconductors from Meissner repulsion}

\author{Lan Luan}
\affiliation{Stanford Institute for Materials and Energy Science, SLAC National Accelerator Laboratory, 2575 Sand Hill Road, Menlo Park, CA 94025}
\author{Ophir M. Auslaender}
\affiliation{Stanford Institute for Materials and Energy Science, SLAC National Accelerator Laboratory, 2575 Sand Hill Road, Menlo Park, CA 94025}
\affiliation{Physics Department, Technion-Israel Institute of Technology, Haifa 32000, Israel}
\author{Nadav Shapira}
\affiliation{Physics Department, Technion-Israel Institute of Technology, Haifa 32000, Israel}
\author{Douglas A. Bonn}
\author{Ruixing Liang}
\author{Walter N. Hardy}
\affiliation{Department of Physics and Astronomy, University of British Columbia, Vancouver, BC, Canada V6T 1Z1}
\author{Kathryn A. Moler}
\affiliation{Stanford Institute for Materials and Energy Science, SLAC National Accelerator Laboratory, 2575 Sand Hill Road, Menlo Park, CA 94025}


\begin{abstract}
We report a method to locally measure the penetration depth $\lambda$ in a superconductor by detecting the diamagnetic response using magnetic force microscopy (MFM). We extract $\lambda$ by fitting the height dependence of the levitation force in the Meissner state using an analytical model that approximates the MFM tip as a single-domain, truncated conical shell. We demonstrate on two \YBCO\ single crystals with two MFM tips that the obtained values agree well with previous results. This approach is not affected by the tip width and can be applied to similar but not identical tips. 
\end{abstract}


\maketitle

The magnetic penetration depth $\lambda$, determined by the density of the superconducting charge carriers, is one of the two fundamental length scales in superconductors \cite{tinkham_introduction_1975}. Accurate determination of $\lambda$ is important for understanding fundamental properties of the superconducting states, such as the order parameter symmetry and the pairing mechanism \cite{hardy_precision_1993, emery_importance_1995}. However, the absolute value of $\lambda$ is notoriously difficult to measure. Established methods, including muon-spin-rotation \cite{Sonier1994}, infrared spectroscopy \cite{Basov_Inplane_1995}, micro-wave cavity techniques \cite{hardy_precision_1993} and lower critical field measurements \cite{Liang05}, average over bulk samples. Spatially resolved techniques may be helpful especially if sample topography, inhomgogeneity, or intrinsic variation are suspected. 

Efforts to measure $\lambda$ by magnetic scanning probes include scanning SQUID susceptometry \cite{hicks_evidence_2009}, mostly limited by the accuracy in determining the sensor-sample separation $z$, and magnetic force microscopy (MFM) by imaging individual vortices \cite{Roseman_MFM_2001, nazaretski_direct_2009}. The quantitative determination of $\lambda$ from vortex images is highly non-trivial due to the convolution with the tip structure, which requires numerically calculating the convoluted signal based on detailed knowledge of the tip magnetic structure \cite{nazaretski_direct_2009}.

In this letter, we report MFM measurements of the absolute value of $\lambda$ by measuring and modeling the height dependence of the diamagnetic response in the Meissner state. The diamagnetic response is not as badly affected by the finite tip width as lateral imaging, allowing us to make approximations that give an analytical description justified for similar tips. We demonstrate on two \YBCO\ (YBCO) single crystals that the values we obtained with two different tips both agree well with previously published results. 

The measurements were performed in a home-built variable-temperature MFM apparatus. We use high resolution cantilevers (NC-18 from Mikro-Masch), and coat the tip by electron beam deposition with an iron film of nominal thickness 40~nm. We measure the change of the resonant frequency \cite{Albrecht_1991}, which is proportional to $\partial F_z/\partial z$, where $\hat{z}$ is normal to the cantilever and to the crystal \textit{a-b} surface, and $\vec{F}$ is the force that results from integrating over the entire tip. The YBCO single crystals were grown by the self-flux method in BaZrO$_3$ crucibles \cite{Liang1998} and annealed, with superconducting transition temperature $T_c\approx 56$~K, implying $x\approx0.56$. The samples are platelet shaped, with the face parallel to crystal $ab$-plane about $1{\rm mm}\times0.7{\rm mm}$ and thickness 60-80 $\mu$m. The samples were kept for less than two months after growth either in a room temperature desiccator or below 77~K ensuring that they were fresh. Sample I is fully detwinned, and sample II has twin boundaries separated by a few microns. 

\begin{figure}
\includegraphics[width=3.3in]{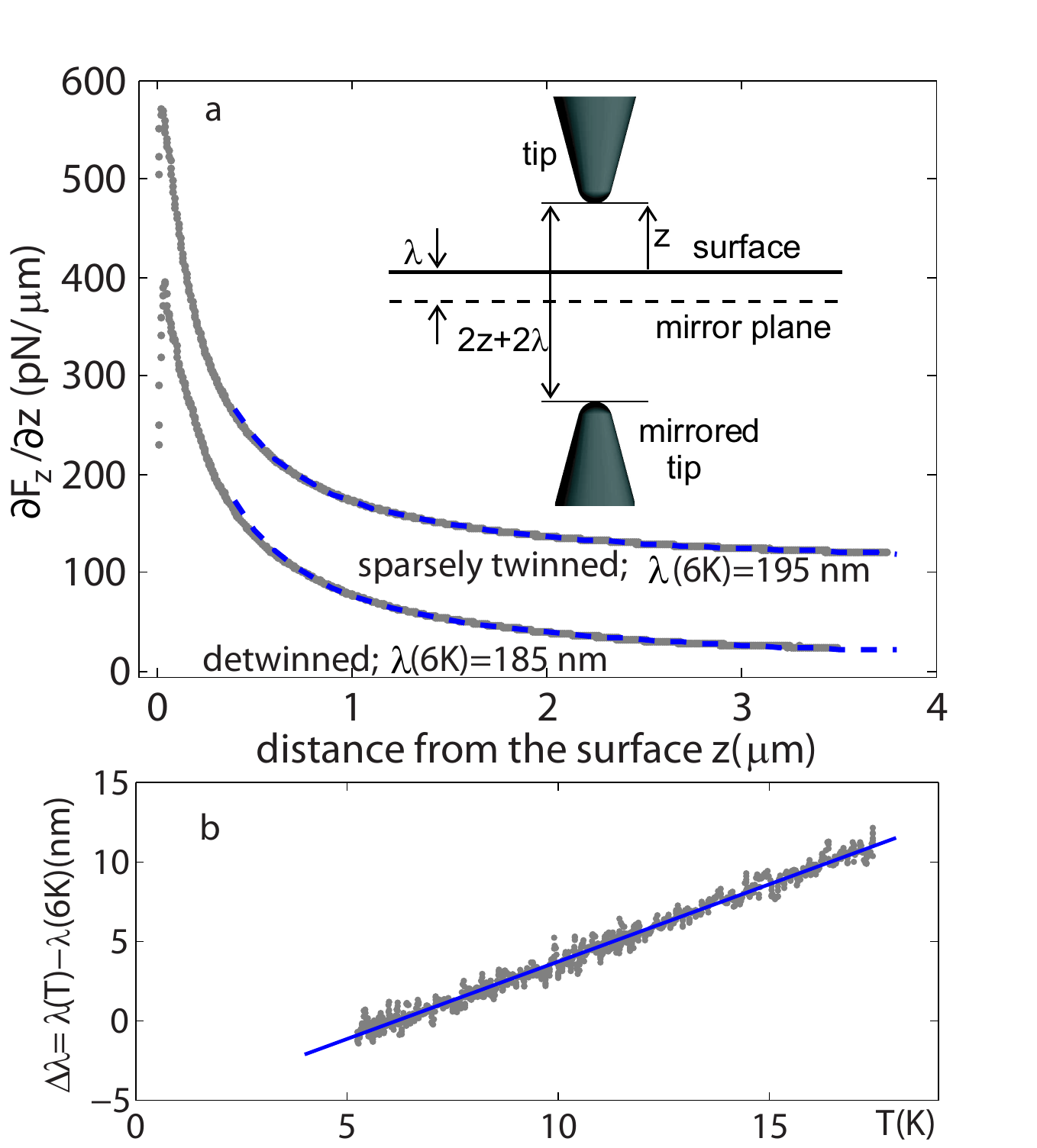}
\caption{\label{fig_TD_sketch} Measuring $\lambda$ and  $\Delta\lambda$ by MFM in the Meissner state using time-reversed mirror approximation. %
{\bf a:} $z$ dependence of $\partial F_z/\partial z$ (grey dots) by MFM at $T=6$~K on two YBCO single crystals ($T_c\approx56$~K) and the fit to the truncated cone model (dashed line) from which we extract $\lambda$ as a fit parameter. The vertical offset, $\partial F_z/\partial z|_{z=\infty}$ is 0 and 100 pN$/\mu$m respectively. Inset illustrates the time-reversed mirror approximation: the response of the superconductor can be replaced by an image tip reflected over a plane $\lambda$ below the superconducting surface. %
{\bf b:} $\Delta\lambda(T)\equiv\lambda(T)-\lambda(6K)$ of sample II (grey dots) from T=6~K to 12~K determined independently of the tip model. The linear fit (solid line) shows a slope of 0.97 nm$/$K. %
}
\end{figure}

To obtain $\lambda$, we cool the samples in the absence of magnetic field and measure $\partial F_z/\partial z(z)$ (Fig.~\ref{fig_TD_sketch}(a)).

The Meissner levitation force on a magnet from a superconductor occupying half space was given in Eq.(2.18) of ref~\cite{xu_magnetic_1995}, from which we obtain the force derivative:
\begin{eqnarray}\label{eq_tip_integral}\nonumber
  \hspace{-0.5cm}  \frac{\partial F_z}{\partial z}(z) &=& -\frac{\mu_0}{2\pi}\int_0^\infty\!\!\!dk k^4 G(\lambda k)e^{-2zk}\int_{\rm tip}\!\!\!\!d{\bf r'}\int_{\rm tip}\!\!\!\!d{\bf r''} \\ 
&& \hspace{-1cm}M({\bf r'})M({\bf r''})e^{-k(z'+z'')}J_0(k|{\bf R'-R''}|) \label{eq_dFzdz}
\end{eqnarray}
where ${\bf r}={\bf R}+z{\hat z}$, ${\bf R}=R{\hat R}$, $J_0$ is the zeroth order Bessel function of the first kind, and for $\lambda k\ll 1$
\begin{equation}\label{eq_G}
G(\lambda k)=\frac{\sqrt{1+(\lambda k)^2}-\lambda k}{\sqrt{1+(\lambda k)^2}+\lambda k}\sim e^{-2\lambda k}\left(1+O[\lambda k]^3\right)
\end{equation}

Approximating $G(\lambda k)$ by $e^{-2\lambda k}$ is equivalent to replacing the response of the superconductor by an image of the field source mirrored through a plane $\lambda$ below the superconducting surface (illustrated in Fig.~\ref{fig_TD_sketch}(a)). Under this time-reversed mirror approximation, the change in $\lambda$ is nearly equivalent to change in $z$, allowing us to determine $\Delta\lambda(T)\equiv\lambda(T)-\lambda(6K)$ independent of any model of the sensor structure \cite{hicks_evidence_2009, Luan_local_2010}. We obtain from sample II a linear $\Delta\lambda(T)$ (Fig.~\ref{fig_TD_sketch}(b)) with a slope consistent with previous results \cite{Panagopoulos_PRL81}, as expected from the nodal d-wave gap structure of YBCO \cite{hardy_precision_1993}.

To extract $\lambda$, we model the tip as a sharp, single-domain conical shell truncated at the distance $h_0$. $M({\bf r})=M_0t\delta(R-\alpha(z+h_0))$, where we assume $\vec{M}$ is along $\hat{z}$,  $\alpha\approx15^o$ is the cone angle, $t$ is the magnetic film thickness, $M_0$ is the magnetization, and $h_0$ is the truncation height. We take the approximation of $J_0(k|{\bf R-R'}|)\approx J_0(0)$ in the limit of $\alpha\ll1$ and large $z$. From Eq.~\ref{eq_tip_integral}, we obtain:
\begin{eqnarray}\label{eq_dFz_trun} \nonumber
\hspace{-0.4cm} \partial F_z/\partial z(z) &=& -\frac{\mu_0}{2\pi}(\alpha M_0 t)^2\int_0^\infty\!\!\!\!dk k^4 G(k\lambda)e^{-2zk}\times\\ 
&& \hspace{-1cm}\left(\int_{\theta_1}^{\theta_2}\!\!d\theta'\int_{0}^\infty\!\!dz' (z'+h_0)e^{-kz'}\right)^2 
\end{eqnarray}

Taking the time-reversed mirror approximation in Eq.\ref{eq_G}, we obtain:
\begin{equation}\label{eq_truncated}\nonumber
\frac{\partial F_z}{\partial z}(z)-\left.\frac{\partial F_z}{\partial z}\right|_{z=\infty} = A\left[\frac{1}{z+\lambda}+\frac{h_0}{z+\lambda}+\frac{ h_0^2}{2(z+\lambda)^2}\right]
\end{equation}
where $A\equiv-\mu_0(\alpha M_0 t\Delta\theta)^2/2\pi$ and $\Delta\theta\equiv\theta_2-\theta_1=\pi$ for our half-coated tips. Since $M_0$ and $t$ may not be known exactly we take $A$ as a fitting parameter. 

\begin{figure}
\includegraphics[width=3.3in]{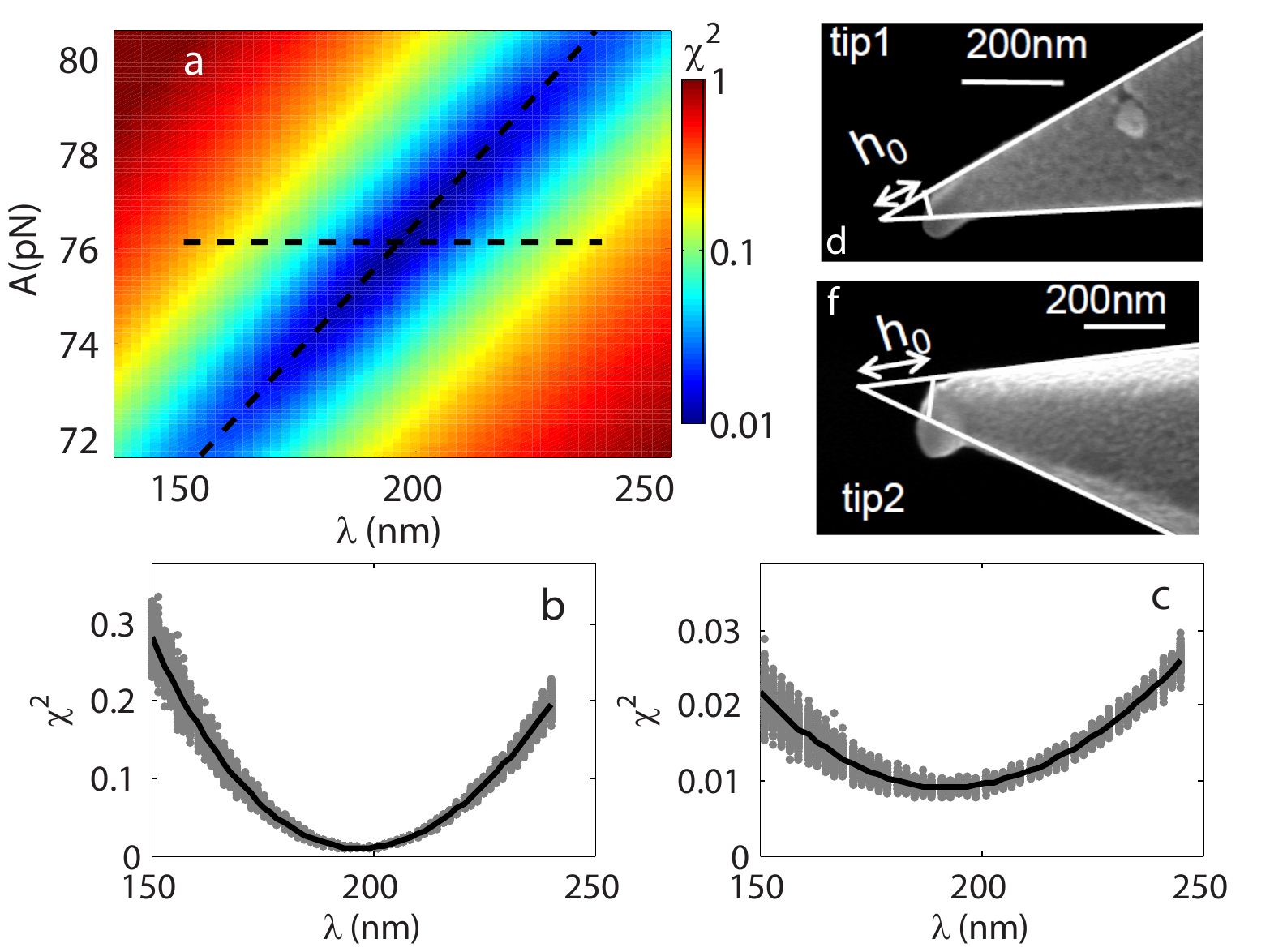}
\caption{\label{fig_tip_error} %
{\bf a-c:} $\chi^2$ for the fit in Fig.~\ref{fig_TD_sketch}(a) for sample II. {\bf a:} $\chi^2$ plotted as the color scale in the two-dimensional parameter space of $A$ and $\lambda$ with $\partial F_z/\partial z$ optimized at each point.  {\bf b-c:} $\chi^2$ as a function of $\lambda$ when $A$ is fixed {\bf (b)} at the value that minimizes $\chi^2$ as plotted by the horizontal dashed line in (a), and when $A$ is free floating {\bf (c)} as given by the diagonal dashed line in (a). The distribution of $\chi^2$ (grey dots) are from bootstrapping 200 times. %
{\bf d-f:} Scanning electron microscopy images of the tip used for sample I (d) and for sample II (f), from which we determine $h_0=120\pm20$~nm and $160\pm20$~nm respectively. The curvature at the tip apex comes from extra deposition of materials due to a sharp edge. %
}
\end{figure}

We fit touchdown curves in Fig.~\ref{fig_TD_sketch}(a) using Eq.~\ref{eq_truncated}, fixing $h_0$ from scanning electron microscopy (SEM) images of the tip (Fig.~\ref{fig_tip_error}(d-f)) and letting $A$, $\lambda$ and $\partial F_z/\partial z|_{z=\infty}$ vary. In the fit, we minimize $\chi^2\equiv \sum (1-\partial F_z/\partial z(z)_{data}/\partial F_z/\partial z(z)_{fit})^2$ for $z\ge0.6\mu$m (Fig.~\ref{fig_tip_error}). We obtain $\lambda(6K)=185$~nm and 196~nm in sample I and sample II respectively. Extrapolating using $\Delta\lambda(T)/T=0.97$ nm$/$K (Fig.~\ref{fig_TD_sketch}(b)), we obtain $\lambda(0)=179$ and 190~nm, in good agreement with previous reported values $\lambda(0)=180\pm20$ nm on similar crystals \cite{Liang2006, Homes1999, Pereg-Barnea_YBCO_2004, Sonier_YBCO_1997}summaries in Table~\ref{TB_lambda}. 
\begin{table} 
\caption{$\lambda$ of \YBCO\ ($x\approx 0.56$) showing good agreement between our results and previously published results. The technique, the obtained value of $\lambda$, the measurement temperature $T_{m}$, and $T_c$ are provided. $\lambda_{ab}$ denotes the the average of $\lambda$ along crystal $a$ and $b$ axes ($\lambda_{ab}\equiv\sqrt{\lambda_a\lambda_b}$). \label{TB_lambda}}
\begin{tabular}{ccccc} \hline \hline 
method & $\lambda$ [nm] & $T_{m}$ [K] & $T_c$ [K] & Ref. \\
\hline $\mu$sR & $\lambda_{ab}=175$ & $1.25$ & $59$ & ref.~\cite{Sonier_YBCO_1997} \\
lower critical field & $\lambda_{ab}=175\pm6$ & 0 & $56$ & ref.~\cite{Liang05} \\
 ESR  & $\lambda_a=202\pm22$ & 0 & $56$ & ref.~\cite{Pereg-Barnea_YBCO_2004} \\
 (Gd-doped) & $\lambda_b=140\pm28$ &  & &  \\
Infrared  &  $\lambda_a=248$ & 12 & $59$ & ref.~\cite{Homes1999} \\
spectroscopy & $\lambda_b=183$ &  &  &   \\
MFM & $\lambda_{ab}=195\pm28$ & 6 & 56 & \\
MFM & $\lambda_{ab}=180\pm30$ & 6 & 56 & \\ 
\hline \hline
\end{tabular}
\end{table}

How accurate is the measurement of $\lambda$? Using the measurement on sample II as an example, if we consider only statistical errors, we obtain $\lambda_{ab}(6K)=196\pm3$~nm from bootstrapping \cite{Efron_Bootstrap_1993} with 70\% confidence interval. 

The systematic errors mainly come from the uncertainty in determining $h_0$ and $z$, and the approximation made in the model. The $\pm 20$ nm uncertainty on $h_0$ leads to $\pm14$ nm uncertainty in $\lambda$. We detect the surface within $\pm$5~nm owing to the abrupt change of $\partial F_z/\partial z$ when van der Waals force dominates over magnetic force (Fig.~\ref{fig_TD_sketch}(a)). We calibrate the scanner using the laser interference pattern from the sample, which gives at most 3\% error in determining $z$ mostly from the nonlinearity of the scanner. We choose to measure fresh surfaces because any non-superconducting layer makes the measured $\lambda$ larger than the real value by the thickness of the layer.  

We observe no change in the tip magnetic strength before and after ramping the magnetic field to 0.1~T at 30~K, consistent with the tip being mono-domain. The magnetization of the tip is presumably aligned along the film. Assuming $\vec{M}$ along $z$ does not induce systematic error in $\lambda$ because the in-plane component gives the same functional dependence of $\partial F_z/\partial z$ as the z component. 

The model ignores the tip width. To estimate the associated error, we release this approximation by taking the Bessel function in Eq.~\ref{eq_dFzdz} to the second order: $J_0(x)=1-1/4x^2+O(x^4)$ and obtain:  
\begin{eqnarray*}
\hspace{-0.3cm}   \frac{\partial F_z}{\partial z}(z)&=&-\frac{\mu_0}{2\pi}(\alpha M_0 t)^2\int_0^\infty\!\!\!\!dk k^4 e^{-2k(\lambda+z)}\int\!\!d\theta' d\theta'' dz' dz'' \\
    &&\hspace{-1.6cm}z'z''e^{-k(z'+z'')}\left(1-\frac{1}{4}\alpha^2 k^2(z'^2+z''^2-2z'z''\cos{(\theta'-\theta'')})\right)
\end{eqnarray*}

Carrying out the integral, the correction on $\partial F_z/\partial z$ is: 
\begin{eqnarray}\label{Eq_tipwidth_corr}\nonumber
\frac{\partial F_z}{\partial z}(z)_{corr} &\approx & A\alpha^2\left(\frac{1}{z+\lambda}+\frac{h_0}{(z+\lambda)^2}+\frac{h_0^2}{2(z+\lambda)^3}\right) \times\\
&& \hspace{-1.3cm} (\frac{3}{2} + \frac{4}{\pi^2}) + A\alpha^2\left(\frac{3}{8}+\frac{2}{\pi^2}\right)\frac{h_0^2}{(z+\lambda)^3}
\end{eqnarray}  

The first term on the right hand side of Eq.~\ref{Eq_tipwidth_corr} has the same functional form as the model in Eq.~\ref{eq_truncated}, so only the second term leads to errors in $\lambda$, at most 2\% owing to the smallness of $\alpha$. Numerical simulations of a truncated cone tip with realistic width quantitatively confirms the analytical model of Eq.~\ref{eq_truncated}. If the tip is a sharp cone, e.g. $h_0=0$, the finite tip width does not change the functional form of $\partial F_z/\partial z$ at all. 

Adding all the errors, we obtain $\lambda(0)=190\pm28$ nm for sample II, and for sample I $\lambda(0)=179\pm30$ nm.

In YBCO crystals, $\lambda$ along the crystal $a$ and $b$ axes are not equal, and we measure the average  $\lambda_{ab}\equiv\sqrt{\lambda_a\lambda_b}$. When pinning is weak, we can determine $\lambda_a/\lambda_b$ by resolving the positions of individual field-cooled vortices. We demonstrated previously that the Fourier transform of vortex positions in a fully doped YBa$_2$Cu$_3$O$_{7-\delta}$ shows an elliptic band, the eccentricity of which gives $\lambda_a/\lambda_b$ \cite{Auslaender08}. However, the weak pinning condition may often not be satisfied \cite{Luan_local_2010}. 

To conclude, we demonstrate a method to measure $\lambda$ within 15\% error by MFM from the height dependence of the diamagnetic response. The essential elements of the method are the precise height determination, and modeling the diamagnetic response with a small number of parameters. 

Acknowledgement: This work was supported by US DoE, Office of Basic Energy and Sciences under Contract No. DEAC02-76SF00515, the Natural Science and Engineering Research Council of Canada and the Canadian Institute for Advanced Research, and the Posnansky Research Fund for High Temperature Superconductivity. We thank Anand Natarajan for his assistance on numerical simulation. 
\providecommand{\noopsort}[1]{}\providecommand{\singleletter}[1]{#1}%

\end{document}